\documentclass{svproc}
\usepackage{url}
\usepackage{hyperref}
\hypersetup{colorlinks,%
citecolor=black,%
filecolor=black,%
linkcolor=red,%
urlcolor=blue,%
pdftex}

\usepackage{amsmath,amsfonts}
\newcommand{\bqn}{\begin{eqnarray}}
\newcommand{\eqn}{\end{eqnarray}}
\newcommand{\bq}{\begin{eqnarray*}}
\newcommand{\eq}{\end{eqnarray*}}

\usepackage{graphicx}
\usepackage{epstopdf}
\usepackage{color}

\begin{document}
\mainmatter              
\title{Embedding of Functional Human Brain Networks on a Sphere}
\titlerunning{Embedding of Functional Human Brain Networks}  
%
\author{Moo K. Chung \and Zijian Chen}
\authorrunning{Moo K. Chung and Zijian Chen} 

\institute{University of Wisconsin, Madison, WI 53706 USA}
\maketitle              

\begin{abstract}
Human brain activity is often measured using the blood-oxygen-level dependent (BOLD) signals obtained through functional magnetic resonance imaging (fMRI). The strength of connectivity between brain regions is then measured as a Pearson correlation matrix. As the number of brain regions increases,  the dimension of matrix increases. It becomes extremely cumbersome to even visualize and quantify such weighted complete networks. To remedy the problem, we propose to embed brain networks onto a sphere, which is a Riemannian manifold with constant positive curvature. The Matlab code for the spherical embedding is given in \url{https://github.com/laplcebeltrami/sphericalMDS}.
 \end{abstract}

\section{Introduction}
\label{sec:Introduction}

\begin{figure}[t]
\begin{center}
\includegraphics[width=0.95\linewidth]{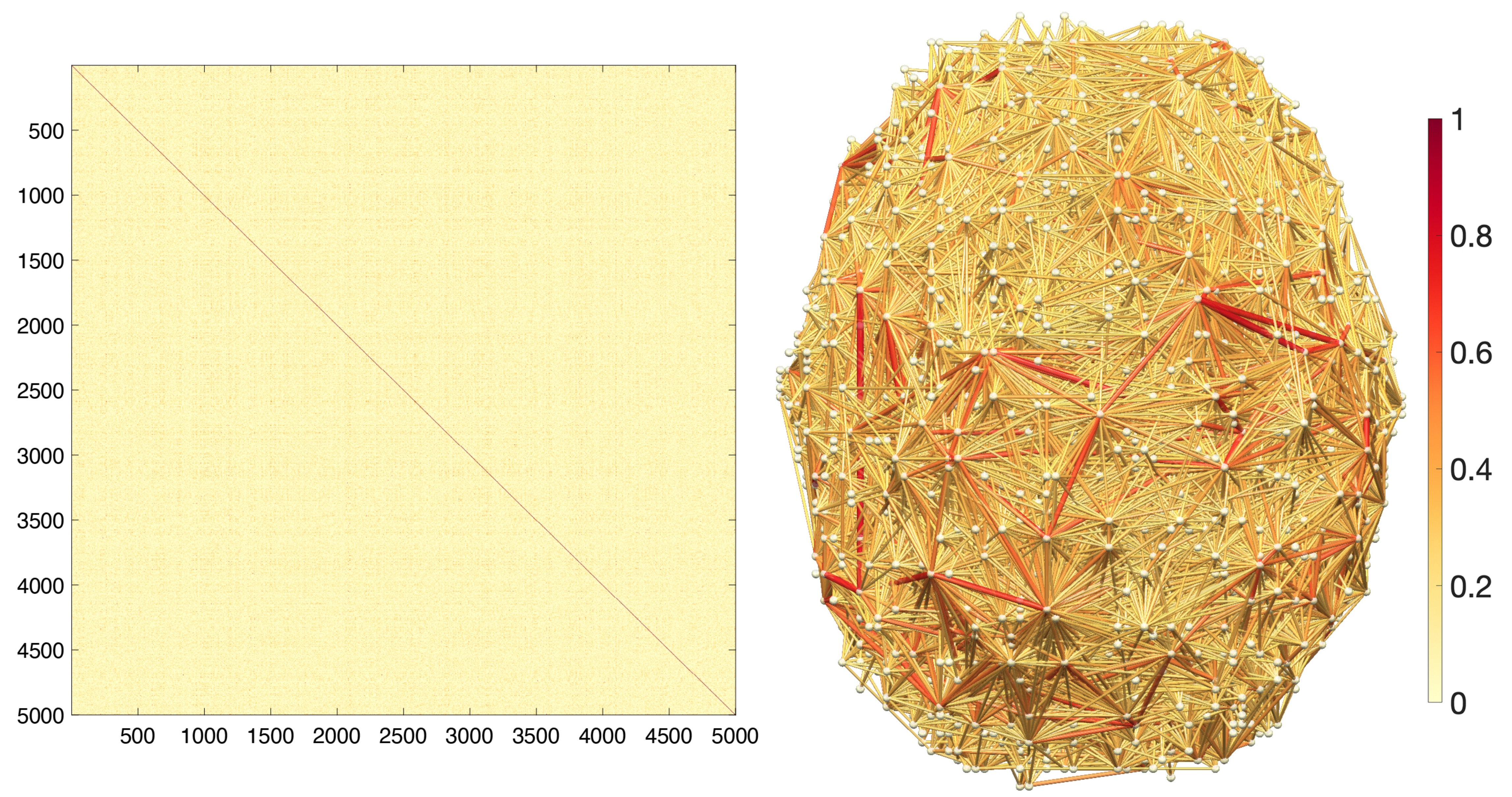}
\caption{The correlation brain network (left) obtained from the resting-state functional magnetic resonance images (rs-fMRI) of a healthy subject at 5000 regions \cite{chung.2018.SPL,gritsenko.2020}. The network is so dense to visualize all the edges (right), only the connections with positive correlations are shown.  Beyond visualization, performing various matrix and graph operations on such a dense network is computationally time consuming and often not feasible.}
\label{fig:persist-cycle}
\end{center}
\end{figure}

In functional magnetic resonance imaging (fMRI) studies of human brain, brain activity is measured by the  blood-oxygen-level dependent (BOLD) contrast. This can be used to map neural activity in the brain. The strength of neural connectivity between regions is then measured using the Pearson correlation (Figure \ref{fig:persist-cycle})  \cite{chung.2019.NN}.  The whole brain is often parcellated into $p$ disjoint regions, where $p$ is  usually a few hundreds \cite{arslan.2018,glasser.2016}.  Subsequently, either functional or structural information is overlaid on top of the parcellation and $p \times p$ connectivity matrix $C = (c_{ij})$ that measures the strength of connectivity between brain regions $i$ and $j$ is obtained. Recently, we are beginning to see large-scale brain networks that are more effective in localizing regions and increasing prediction power \cite{chung.2018.SPL,valencia.2009}. However, increasing parcellation resolution also increases the computational burden exponentially.

For an undirected network with $p$ number of nodes, there are $p(p-1)/2$ number of edges and thus, the brain network is considered as an object in dimension $p(p-1)/2$. Such high dimensional data often requires $\mathcal{O}(p^3)$ run time for various matrix manipulations such as matrix inversion and singular value decomposition (SVD). Even at 3mm resolution, fMRI has more than 25000 voxels in the brain \cite{chung.2018.SPL}. It requires about 5GB of storage to store the matrix of size $25000 \times 25000$ (Figure \ref{fig:persist-cycle}). At 1mm resolution, there are more than 300000 voxels and it requires more than 670GB of storage. Considering large-scale brain imaging datasets such has HCP (Human Connetome Project \url{http://www.humanconnectomeproject.org}) often have thousands images, various learning and inference at higher spatial resolution would be a serious computational and storage challenge. We directly address these challenges by embedding the brain networks into 3D unit sphere. Although there are few available technique for embedding graphs into hyperbolic spaces, the hyperbolic spaces are not intuitive and not necessarily easier to compute \cite{munzner.1998,shi.2019}. Since numerous computational tools have been developed for spherical data, we propose to embed brain networks on a 3D sphere and utilizes such tools. The spherical network embedding will offer far more adaptability and applicability in various problems.

\section{Embedding onto a sphere}

Consider a weighted complete graph $G=(V, C)$ consisting of node set $V = \left\{ 1, \dots, p \right\}$ and edge weight  $C=(c_{ij})$, where $c_{ij}$ is the weight between nodes $i$ and $j$. The edge weights in most brain networks are  usually given by some sort of similarity measure between nodes \cite{lee.2011.MICCAI,li.2009,mcintosh.1994,newman.1999,song.2005}. Most often the Pearson correlation is used in brain network modeling \cite{chung.2019.NN}  (Figure \ref{fig:persist-cycle}).

Suppose measurement vector ${\bf x}_j = (x_{1j},\cdots,$ $x_{nj})^\top \in \mathbb{R}^n$ is given on node $j$ over $n$ variables or subjects. We center and rescale the measurement ${\bf x}_j$ such that  
$$ {\bf x}_{j}^{\top}{\bf x}_{j} = \sum_{i=1}^n x_{ij}^2 = 1, \quad \sum_{i=1}^n x_{ij} = 0.$$
We can show that $c_{ij} = {\bf x}_i^\top{\bf x}_j$ is the Pearson correlation \cite{chung.2019.NN}.
Such points are in the $n$-dimensional sphere $S^{n-1}$. Let the  $n \times p$ data matrix be
${\bf X} = [{\bf x}_1, \cdots, {\bf x}_p].$
Then the correlation matrix is given by ${\bf X}^{\top}{\bf X}$
\bqn {\bf X}^{\top}{\bf X} =  ({\bf x}_i^{\top} {\bf x}_j) = U^{\top} D(\eta_1, \cdots, \eta_p)  U \label{eq:SVD}\eqn
with $U^{\top}U = I_p$ and $D$ is a diagonal matrix with eigenvalues $\eta_1 \geq \cdots \geq \eta_p$. Since there are usually more nodes than variables in brain images, i.e., $n \ll p$, the correlation matrix is not positive definite and some eigenvalues might be zeros \cite{chung.2018.SPL}. Since the correlation is a similarity measure, it is not distance. 
Often used correlation distance 
$c_{ij} = 1- {\bf x}_i^\top{\bf x}_j$ is not metric. To see this, consider the following 3-node counter example:
\bq {\bf x}_1 = (0, \frac{1}{\sqrt{2}},  -\frac{1}{\sqrt{2}})^\top, \quad {\bf x}_2 = (\frac{1}{\sqrt{2}},  0, -\frac{1}{\sqrt{2}})^\top, \quad {\bf x}_3 &=& (\frac{1}{\sqrt{6}}, \frac{1}{\sqrt{6}}, -\frac{2}{\sqrt{6}})^\top.
\eq
Then we have $c_{12} > c_{13} + c_{23}$ disproving triangle inequality. Then the question is under what condition, the Pearson correlations becomes a metric?

\begin{theorem}For centered and scaled data ${\bf x}_1, \cdots, {\bf x}_p$, 
$\theta_{ij} = cos^{-1}({\bf x}_i^\top{\bf x}_j)$ is a metric. 
\label{theorem:metric}
\end{theorem}
{\em Proof.}
The centered and scaled data  ${\bf x}_i$ and ${\bf x}_j$ are  residing on the unit sphere $S^{n-1}$.  The correlation between ${\bf x}_i$ and ${\bf x}_j$ is the cosine angle $\theta_{ij}$ between the two vectors, i.e., $${\bf x}_i^\top {\bf x}_j = \cos \theta_{ij}.$$ 
The geodesic distance $d$ between nodes ${\bf x}_i$ and ${\bf x}_j$ on the unit sphere is given by angle $\theta_{ij}$:
$$\theta_{ij} = \cos^{-1} ({\bf x}_i^\top {\bf x}_j).$$ 
For nodes ${\bf x}_i, {\bf x}_j \in S^{n-1}$, there are two possible angles $\theta_{ij}$ and $2\pi - \theta_{ij}$ depending on if we measure the angle along the shortest or longest arcs. We take the convention of using the smallest angle in defining $\theta_{ij}$. 
With this convention, 
$$\theta_{ij}  \leq \pi.$$
Given three nodes ${\bf x}_i, {\bf x}_j$ and ${\bf x}_k$, which forms a spherical triangle, 
we then have spherical triangle inequality
\bqn \theta_{ij} \leq \theta_{ik} + \theta_{kj}. \label{eq:STI} \eqn
The inequality (\ref{eq:STI}) is proved in \cite{reid.2005}. Other conditions for metric are trivial. \hfill $\qed$

Theorem \ref{theorem:metric} shows that measurement vector ${\bf x}_j$ can be embedded onto a sphere $S^{n-1}$ without any loss of information. The geodesic $d({\bf x}_i, {\bf x}_j)$ on the sphere is the Pearson correlation. A similar approach is proposed for embedding an arbitrary distance matrix to a sphere in \cite{wilson.2014}. In our case, the problem is further simplified due to the geometric structure of correlation matrices \cite{chung.2019.NN}.

\section{Hypersperical harmonic expansion of brain networks}
Once we embedded correlation networks onto a sphere, it is possible to algebraically represent such networks as basis expansion involving the hyperspherical harmonics \cite{domokos.1967,higuchi.1987,hosseinbor.2015.MIA2,hosseinbor.2015.MIA1,hosseinbor.2014.MICCAI}.
Let $\boldsymbol{\varphi} = (\theta, \varphi_2, \cdots, \varphi_{n-1})$ be the spherical coordinates of $S^{n-1}$ such that $\theta \in [0,\pi), \varphi_i \in [0,2\pi)$ where $\theta$ is the axial angle.  Then the spherical Laplacian $\Delta_{n-1}$ is iteratively given as \cite{cohl.2011}
$$\Delta_{n-1}  = \frac{\partial^2}{\partial \theta^2} + (n-2) \cot \theta \frac{\partial}{\partial \theta}+ \frac{1}{\sin^2 \theta} \Delta_{n-2}.$$
With respect to the spherical Laplacian $\Delta_{n-1}$, the hyperspherical harmonics $Y_{\bf l} (\boldsymbol{\varphi} )$ with ${\bf l} =  (l_1, \cdots, l_{n-1})$ satisfies
$$ \Delta_{n-1} Y_{\bf l}(\boldsymbol{\varphi} ) = - \lambda_{\bf l} Y_{\bf l}(\boldsymbol{\varphi} )$$
with eigenvalues $\lambda_{\bf l} =  l_{n-1}(l_{n-1} + n-1)$ for $|l_1 | \leq l_2 \leq \cdots l_{n-1}$. The hyperspherical harmonics are given in terms of the Legendre polynomials. 
We can compute the hyperspherical harmonics inductively from the previous dimension starting with $S^2$, which we parameterize with  $(\theta, \varphi_2) \in [0,\pi) \otimes [0,2\pi)$, where  the spherical harmonics $Y_{l_2 l_1}$ are given by \cite{chung.2008.sinica,courant.1953}
\bq
Y_{l_2 l_1} = 
\begin{cases} 
c_{l_2 l_1}P^{|l_1|}_{l_2}(\cos\theta)\sin
(|l_1|\varphi_2), &-l_2 \leq l_1\leq -1, \\
 \frac{c_{l_2 l_1}}{\sqrt{2}}P_{l_2}^{| l_1|}(\cos\theta),& L_1=0,\\
c_{l_2 l_1} P^{| l_1 |
}_{l_2}(\cos\theta)\cos (|l_1 |\varphi_2),& 1 \leq l_1\leq l_2,
 \end{cases}
\eq
with normalizing constants $c_{l_2 l_1}$ and 
the associated Legendre polynomial $P_{l_2}^{l_1}(x)$ \cite{chung.2008.sinica}. 

The hyperspherical harmonics are orthonormal with respect to area element
$$d\mu(\boldsymbol{\varphi}) = \sin^{n-2} \theta \sin^{n-3} \varphi_2 \cdots \sin \varphi_{n-1} d \boldsymbol{\varphi}$$
such that 
\bq \langle Y_{{\bf l}_1} Y_{{\bf l}_2} \rangle &=& \int_{S^{n-1}}Y_{{\bf l}_1} (\boldsymbol{\varphi} ) Y_{{\bf l}_2} (\boldsymbol{\varphi} ) \;d\mu(\boldsymbol{\varphi})   = \delta_{{\bf 1}_1 {\bf 1}_2},\eq 
where $\delta_{{\bf 1}_1 {\bf 1}_2}$ is the Kronecker's delta. Then using the hyperspherical harmonics, we can build 
the multiscale representation of networks through the heat kernel expansion \cite{chung.2008.TMI,chung.2008.sinica}
\bq K_{t}(\boldsymbol{\varphi},\boldsymbol{\varphi}') = \sum_{\bf l} e^{-\lambda_{\bf l} t} Y_{\bf l}(\boldsymbol{\varphi} ) Y_{\bf l}(\boldsymbol{\varphi}'), \eq
where the summation is over all possible valid integer values of ${\bf l}$.

Given initial data $g(t = 0, \boldsymbol{\varphi}) = f(\boldsymbol{\varphi})$ on $S^{n-1}$, the solution to diffusion equation
$$\frac{dg}{dt} =  \Delta_{n-1} g$$
at time $t$ is given by
\bq g(t, \boldsymbol{\varphi}) &=& \int_{S^{n-1}} K_{t}(\boldsymbol{\varphi},\boldsymbol{\varphi}')  f(\boldsymbol{\varphi}') \; d\mu(\boldsymbol{\varphi}')\\
&=& \sum_{{\bf l}}  e^{-\lambda_{\bf l} t} f_{\bf l}Y_{\bf l}(\boldsymbol{\varphi})
\eq
with spherical harmonic coefficients $f_{\bf l} =  \langle f, Y_{\bf l} \rangle$. The coefficients are often estimated in the least squares fashion in the spherical harmonic representation (SPHARM) often used in brain cortical shape modeling (\url{https://github.com/laplcebeltrami/weighted-SPHARM})
\cite{chung.2008.sinica,gerig.2001,gu.2004,shen.2006}. The embedded network nodes can be modeled as the Dirac delta function such that
$$f (\boldsymbol{\varphi}) = \frac{1}{p} \sum_{i=1}^p \delta(\boldsymbol{\varphi} - {\bf x}_i).$$
We normalize the expression such that
$\int_{S^{n-1}} f (\boldsymbol{\varphi}) d\mu(\boldsymbol{\varphi}) =1.$
Then we can algebraically show that the solution is given by
$$g(t, \boldsymbol{\varphi}) = \sum_{{\bf l}}   e^{-\lambda_{\bf l} t} Y_{\bf l} (\boldsymbol{\varphi} ) \sum_{i=1}^p  Y_{\bf l}({\bf x}_i).$$

\section{Spherical multidimensional scaling}
We have shown how to embed correlation matrices into $S^{n-1}$ and model them parametrically using the spherical harmonics. In many large scale brain imaging studies, the number of variables $n$ can be thousands. Embedding in such a high dimensional sphere may not be so useful in practice. We propose to embed correlation matrices into two sphere $S^2$, which is much easier to visualize and provides parametric representation through available SPAHRM tools  \cite{chung.2008.TMI}.

Given geodesic $\theta_{ij} = \cos^{-1} ({\bf x}_i^{\top} {\bf x}_j)$ in $S^{n-1}$, we want to find the lower dimensional embedding ${\bf y}_j =(y_{1j}, y_{2j}, y_{3j})^{\top} \in S^2$ satisfying
$$\| {\bf y}_j \|^2 =  {\bf y}_j ^{\top}  {\bf y}_j  = \sum_{i=1}^3 y_{ij}^2 = 1.$$
This is a spherical multidimensional scaling (MDS) often encountered in analyzing spherical data such as earthquakes \cite{dzwinel.2005} and colors in computer vision \cite{maejima.2012}. Given data matrix ${\bf X}$ in (\ref{eq:SVD}), we find ${\bf Y}= [{\bf y}_1, \cdots, {\bf y}_p]$ that minimizes the loss
\bqn \mathcal{L}({\bf X}, {\bf Y}) = \sum_{i,j=1}^p   \big[ \cos^{-1} ({\bf x}_i^{\top} {\bf x}_j) - \cos^{-1} ({\bf y}_i ^{\top} {\bf y}_j) \big]^2. \label{eq:sphericalMDS} \eqn
We propose to solve  spherical MDS  (\ref{eq:sphericalMDS}) without the usual gradient descent on spherical coordinates \cite{maejima.2012}. At the minimum of  (\ref{eq:sphericalMDS}), we can approximate the loss linearly using the Taylor expansion \cite{abramowitz.1988}
$$\cos^{-1}  ({\bf x}_i^{\top} {\bf x}_j) = \frac{\pi}{2} - {\bf x}_i^{\top} {\bf x}_j + \cdots$$
as 
\bqn \mathcal{L}({\bf X}, {\bf Y})  = \sum_{i,j=1}^p   \big[ {\bf x}_i^{\top} {\bf x}_j  - {\bf y}_i ^{\top} {\bf y}_j \big]^2. \label{eq:LXY} \eqn
The loss (\ref{eq:LXY}) can be further written in the matrix form 
\bqn \mathcal{L}({\bf X}, {\bf Y}) =  \mbox{tr} (  {\bf X}^{\top}{\bf X} - {\bf Y}^{\top}{\bf Y})^2  = \|  {\bf X}^{\top}{\bf X} - {\bf Y}^{\top}{\bf Y} \|_F^2 \label{eq:LRAP},
\eqn
with the Frobenius norm $\| \cdot \|_F$. The minimization of linearized loss (\ref{eq:LRAP}) is a {\em low-rank approximation problem}, which can be solved  exactly through the Eckart Young Mirsky theorem \cite{golub.1987} that states
$$  \arg \min_{ {\bf Y}^{\top}{\bf Y} }   \mathcal{L}({\bf X}, {\bf Y}) = U^{\top} D(\eta_1, \eta_2, \eta_3, 0, \cdots, 0)  U,$$
where $U$ is the orthogonal matrix obtained as the SVD of ${\bf X}^{\top}{\bf X}$ in (\ref{eq:SVD}) \cite{dax.2014}. In order to match two given symmetric matrices, we need to align with the principle directions of the the matrices. The matching is then the most optimal  in the Frobenius norm sense. However, the result of the Eckart Young Mirsky theorem does not constrain and and embeded points ${\bf Y}= [{\bf y}_1, \cdots, {\bf y}_p]$ are not necessary constrained on the sphere. Thus, we need to normalize ${\bf Y}$ such that ${\bf y}_j \in S^2$ as follows. 

Let $[{\bf v}_1, \cdots, {\bf v}_p]$ be $3 \times p$ matrix consisting of the first three rows of $D(\sqrt{\eta_1}, \sqrt{\eta_2},\\ \sqrt{\eta_3}, 0, \cdots, 0)  U$. All other rows below are zero. Then each column  vector ${\bf v}_j$ is normalized to be embededed on the unit sphere such that $\frac{{\bf v}_j}{ \| {\bf v}_j \|} \in S^2$. Since
$$ \Big( \frac{{\bf v}_i^{\top} {\bf v}_j}{\|{\bf v}_i  \|  \| {\bf v}_j \|} \Big) = U^{\top} D(\eta_1, \eta_2, \eta_3, 0, \cdots, 0)  U,$$
${\bf Y} = \Big[ \frac{{\bf v}_1}{ \| {\bf v}_1 \|}, \cdots, \frac{{\bf v}_p}{ \| {\bf v}_p \|} \Big]$ solves (\ref{eq:LRAP}) and we claim
\begin{theorem}  
$$  \arg \min_{{\bf y}_j \in S^2} \mathcal{L}({\bf X}, {\bf Y}) =  \Big[ \frac{{\bf v}_1}{ \| {\bf v}_1 \|}, \cdots, \frac{{\bf v}_p}{ \| {\bf v}_p \|} \Big] .$$
\end{theorem}

The embedding is not unique. For any rotation matrix $Q$,  $Q{\bf Y} $ will be another embedding.
In brain imaging applications, where we need to embed multiple brain networks, the rotation matrix $Q$ should not come from individual subject but should come from the fixed average template.

\begin{figure}[t]
\begin{center}
\includegraphics[width=1\linewidth]{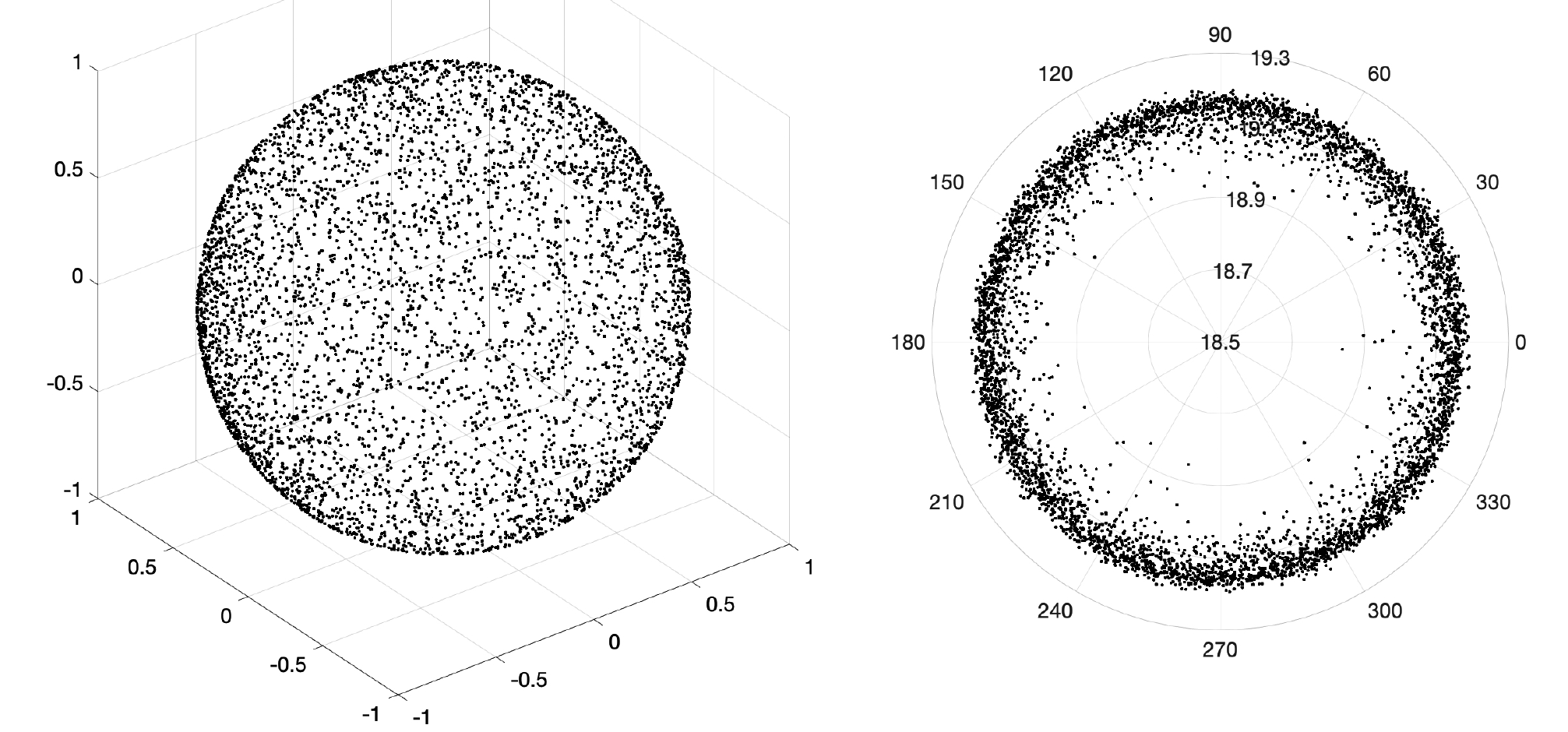}
\caption{Left: embedding of a brain network with 5000 nodes to sphere $S^2$. Each scatter point can be modeled as the Dirac delta function. Right:  embedding to hyperbolic space, where the embedded points from a torus-like circular pattern.}
\label{fig:embedding-sphere}
\end{center}
\end{figure}

\section{Experiement}
\begin{figure}[t]
\begin{center}
\includegraphics[width=1\linewidth]{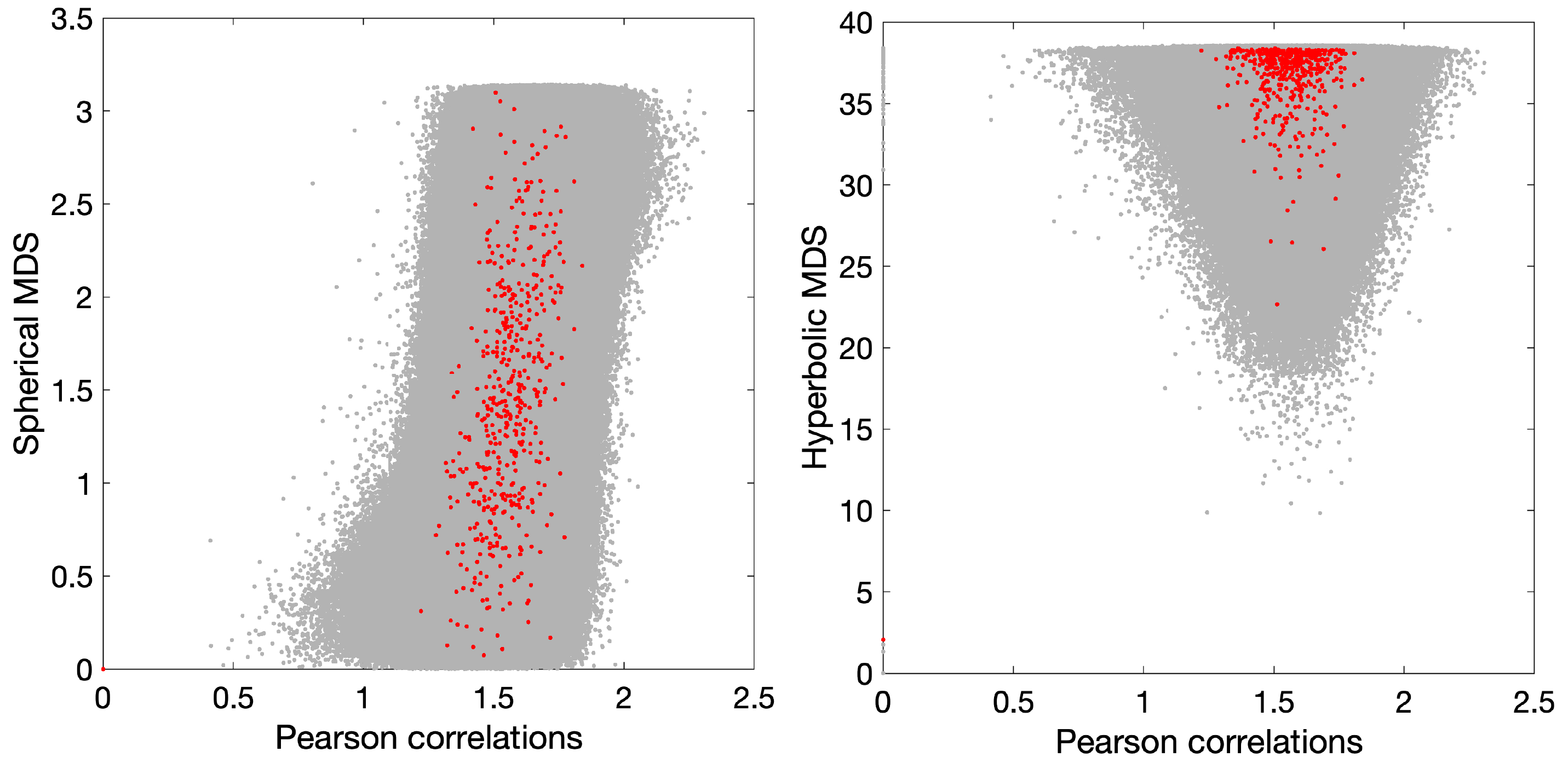}
\caption{The shepard diagrams of all the pairwise distances (gray points) on 5000 nodes with spherical MDS (left) and hyperbolic MDS. Red points are first 500 pairwise distances. The geodesic distance $\cos^{-1} ({\bf x}_i^{\top} {\bf x}_j)$ is used for Pearson correlations. The correlation between distances are 0.51 for spherical MDS and 0.0501 for hyperbolic MDS demonstrating far better embedding performance of spherical MDS.}
\label{fig:dist-dist}
\end{center}
\end{figure}

We applied the proposed spherical embedding to a functional brain network obtained through the Human Connectome Project (HCP) \cite{smith.2013,vanessen.2012}. The resting-state functional magnetic resonance image (rs-fMRI) of a healthy subject was collected with 3T scanner at 2.0 mm voxel resolution (91×109×91 spatial dimensionality), 1200 frames at a rate of one frame every 720 ms. The scan went through spatial and temporal preprocessing including spatial distortions and motion correction, frequency filtering, artifact removal as the part of HCP preprocessing pipeline  \cite{smith.2013}. fMRI is then denoised using the Fourier series expansion with cosine basis \cite{gritsenko.2020}. A correlation matrix of size 5000 $\times$ 5000 was obtained by computing the Pearson correlation of the expansion coefficients  across uniformly sampled 5000 brain regions (Figure \ref{fig:persist-cycle}). Following the proposed method, we embedded the brain network into 5000 scatter points on $S^2$ (Figure \ref{fig:embedding-sphere}). The method seems to embed brain networks uniformly across $S^2$. The Shepard diagram  of displaying distance  $\cos^{-1} ({\bf y}_i^{\top} {\bf y}_j)$  vs.  $\cos^{-1} ({\bf x}_i^{\top} {\bf x}_j)$ is given in Figure \ref{fig:dist-dist}. The correlation between the distances is 0.51 indicating a reasonable embedding performance. The Matlab code for performing the spherical embedding is provided in \url{https://github.com/laplcebeltrami/sphericalMDS}.

\section{Discussion}

The embedding of  brain networks to a sphere allows various computation of brain networks straightforward. 
 Instead of estimating the network gradient discretely in coarse fashion, it is possible to more smoothly estimate 
 the network gradient on a sphere using the spherical coordinates \cite{elad.2005}. We can further able to obtain various differential quantities of brain networks such as gradient and curls often used in the Hodge decomposition \cite{anand.2021.arXiv}. This is left as a future study.

The major body of literature on the embedding of networks is toward  hyperbolic spaces, where 2D Poincare disk $D$ is often used \cite{munzner.1998,shi.2019,wilson.2014}. Figure \ref{fig:embedding-sphere} shows the embedding of the brain network to $D$ using hyperbolic MSD \cite{zhou.2021}. It is usually characterized by the torus-like circular pattern. Unlike spherical embedding, the hyperbolic embedding does not provide uniform scatter points. 

The Shepard diagram  of displaying the geodesic distance in the Poincare disk  vs.  $\cos^{-1} ({\bf x}_i^{\top} {\bf x}_j)$ is given in Figure \ref{fig:dist-dist}. The correlation between distances is 0.0501 indicating a poor embedding performance compared to the spherical MDS. For correlation based brain networks, the spherical MDS might be a better alternative over hyperbolic MDS. However, further investigation is needed for determining the optimal embedding method for brain networks.

\section*{Acknowledgements}
We thank Zhiwei Ma of University of Chicago for providing the 3-nodes counter example. We thank the discussion with Hyekyoung Lee of Seoul National University Hospital on  metrics on correlations. This study is funded by NIH R01 EB022856, EB02875, NSF MDS-2010778.

\bibliographystyle{plain}
\bibliography{reference.2022.03.27}

\end{document}